# Controllable patterning and CVD growth of isolated carbon nanotubes with direct parallel writing of catalyst using Dip Pen Nanolithography

*Irma Kuljanishvili*[1]*, *Dmitriy A. Dikin*[2], *Sergey Rozhok*[3], *Scott Mayle*[1] *and Venkat Chandrasekhar*[1]*

[1]Department of Physics Astronomy, Northwestern University, Evanston, IL, 60208
[2]Department of Mechanical Engineering, Northwestern University, Eavnston, IL, 60208
[3] NanoInk, Inc., Skokie, IL, 60077

*E-mail: i-kuljanishvili@northwestern.edu, v-chandrasekhar@northwesterrn.edu

**Abstract**

We report a process to fabricate carbon nanotubes (CNT) by chemical vapor deposition at predetermined location. This process was enabled by patterning catalyst nanoparticles directly on silicon substrates with nanometer-scale precision using Dip Pen Nanolithography® (DPN®). A multi-pen writing method was employed to increase the patterning rate. The development of new molecular inks for the deposition of the precursor catalyst resulted in a high yield of isolated carbon nanotubes, ideal for subsequent device fabrication. Here, we demonstrate the advantages of the new method for producing high quality isolated CNT in scalable array geometries.

Carbon nanotubes (CNT) are one of the most attractive building blocks for constructing nanoscale devices due to their unique structural, mechanical, electrical, thermal, and optical properties.[1,2] Consequently they have been proposed as components of many devices, including transistors, sensors, logic circuits, interconnects, and components of micro- and nano-electromechanical systems.[3,6]

The chemical vapor deposition (CVD) method using nanoparticle as catalysts deposited on a substrate has proved to be successful for production of high-quality CNT. The size of the catalyst particles is critical in determining whether multi-walled or single-walled CNT are produced. If the size of the catalyst particles is more than a few nanometers in diameter, the yield of single-walled carbon nanotubes (SWNT) decreases.[4,5] Catalyst is often dispersed on the substrate from a solution containing a suspension of the nanoparticles either by spin coating the substrate or simply dipping the substrate into the catalyst solution. Either method will result in a relatively uniform dispersion of nanoparticles on the substrate surface after drying. However, integrating CNT into micro- and nano-devices has proved challenging since they have to be precisely positioned with respect to other device elements. In order to position the catalysts at specific locations, a number of conventional lithographic techniques have been used up to date, but all of these techniques suffer from certain limitations. For example, photolithography does not have the resolution to produce nanometer-scale patterns; electron-beam lithography does have the required resolution, but being a serial patterning process, it can be very time consuming. Furthermore, if conventional lithography techniques are used to deposit the catalyst, depositing nanometer-size catalyst particles on the substrate is not trivial. Microprinting using PDMS stamps has also been widely used, but the stamps themselves have to be patterned anew for each new design.[7,8] Moreover, many of these techniques use a so- called "mask" approach that could potentially introduce contamination on the substrate and negatively impact the growth of nanotubes.

DPN and the related Fountain Pen Nanolithography (FPN) are newer lithography techniques based on scanning probe microscopy. Both have a number of advantages for the fabrication of a variety of nanostructures.[9-12] Generally, for the DPN method employed here, the tip of an atomic force microscope (AFM) is dipped in an "ink" that can subsequently be transferred to a substrate with nanometer-scale precision typical for scanning probe microscopes. The "ink" usually consists of nanoparticles suspended in a liquid, or inorganic or biological molecules in a solvent. DPN has been used for several years to deposit nanoparticles and biological molecules such as DNA and proteins onto surfaces and to pattern metallic surfaces by using chemicals that act as etch masks.[9-12] The pattern is designed in a software program that can be changed readily, and large arrays of identical structures can be fabricated by simultaneously using multiple tips in an ordered array. Millions of identical structures have been patterned in one step using specially fabricated arrays of cantilevers.[13] DPN is a mask free patterning process so that it: 1) is flexible and convenient for modifications of patterned designs, 2) vastly minimizes possible contamination of the substrate, and 3) could be used in a parallel writing application.

DPN consequently provides a number of advantages in patterning catalytic precursors for subsequent CVD growth of carbon nanotubes. However, the reliability and reproducibility of the technique depends critically on the properties of the ink used in the process, and ambient conditions such as humidity and temperature during the patterning process. In addition, the size of the catalyst particles that are formed on the substrate determine the yield of CNT after the CVD process. Recently, a DPN method employing a single AFM tip was used to pattern ink consisting of a colloidal suspension of cobalt nanoparticles for the CVD growth of CNTs.[14] In this work, we demonstrate the use of iron-based molecular type ink delivered by a parallel writing method employing multi-pen cantilevers. We demonstrate the ability to pattern catalyst onto specific predetermined locations with subsequent CVD growth of SWNTs. The molecular ink was developed for this study and optimized to produce high-quality isolated CNT for easy integration into various device architectures. We also demonstrate advantages of applying DPN with multiple pens simultaneously for scalability of this process to very large arrays.

Figure 1 shows a schematic of the entire process that includes patterning, CVD synthesis and characterization of the resulting CNT using a variety of probes. For substrates, commercially available silicon wafers with a thermal oxide layer of thickness 300-400 nm were obtained. An array of metallic fiducial marks was then patterned onto the substrate using conventional electron-beam lithography techniques to aid in alignment and to locate specific nanotubes after CVD growth. Substrates and pen arrays were made hydrophilic prior to the DPN process (Supporting Information SI-2). The desired pattern was then written onto the substrate with an iron-based ink solution (Figure 2a and b). The pattern chosen for this study was a rectangular array of ink dots on the substrate surface. After the ink dried and the substrate was heat treated in an $Ar/H_2$ atmosphere, nanometer-size catalyst particles remained in the patterned areas (Figure 2c). These nanoparticles served as the catalyst for the CVD growth of the CNT. The size of these particles was < 10 nm (see the AFM line profile in Figure 2d), suitable for growing single-walled carbon nanotubes.[4] The lower limit on the size of the patterns that can be produced by the DPN process is determined by a number of factors. These include the composition and viscosity of the ink, the relative humidity and temperature, any treatment of the surface to increase or decrease its hydrophilicity, and any functionalization of the surface to reduce the diffusion of the molecules or nanoparticles once they are deposited on the surface. For growing nanotubes from nanoparticle catalysts, it is critical that the catalyst not be poisoned during the patterning process, since poisoning of the catalyst will reduce the yield of CNT. In this paper, we demonstrate that iron salts conventionally used to produce high-quality SWNT form an excellent basis for the ink used in the DPN process. The salts do not form a suspension or a colloid, but are completely dissolved in the solvent, and the catalyst nanoparticles are created on the surface after evaporation of the solvent and reduction in the CVD furnace. In the simplest case, the solvent is just water. However, the properties of the ink can be varied by using mixtures of solvents to adjust the viscosity, volatility and other properties of the ink, giving one control over the DPN writing process. The two iron-based salts that were used in this study were ferric nitrate nonahydrate [$Fe(NO_3)_3 \cdot 9H_2O$] and ferric chloride hexahydrate [$FeCl_3 \cdot 6H_2O$]. These two salts were used to prepare a master solution with a solute concentration ranging from 0.175 to 0.200 mg/mL in deionized water (18 MΩ-cm). While the solubility of both salts in water is much higher, this range of concentrations was chosen because it gave a suitable density of CNT after CVD growth, as verified by control experiments. Both the ferric chloride solution and the ferric nitrate solution gave good results. Here, we shall concentrate primarily on results from the ferric nitrate solution. The master solution had to be diluted further with organic solvents that were miscible with water, but which had lower vapor pressures under ambient conditions. This additional dilution helped to avoid excess evaporation of the ink during the long writing process (Supporting Information SI-2). It was critical that the addition of these organic solvents did not affect the activity of the catalyst during CVD growth. Best results were achieved by diluting the master solution with a mixture of dimethylformamide (DMF) and glycerol. Figure 3a shows a network of SWNT grown on an oxidized silicon substrate using CVD after dipping the substrate into an ink solution consisting of a mixture of master solution, deionized water, DMF and glycerol (6:2:3:1 parts by volume respectively). [16] The high density of resulting nanotubes shows that the catalyst retained its activity during the CVD growth. The quality of the nanotubes was analyzed by measuring the Raman spectrum, which is illustrated in Figure 3b for three randomly chosen spots on the sample shown in Figure 3a. A characteristic signature of the Raman spectrum of single-walled CNT is the presence of a double peak in the Raman shift at ~1600 cm$^{-1}$. This double peak consists of a smaller so-called *G-* peak at lower wavenumbers, and a larger *G+* peak at higher wavenumbers. The splitting of the *G* peak cannot be seen in Figure 3b, but it can be clearly observed on an expanded scale (see Supporting Information, SI-5, Figure S4). For the disordered SWNT, the Raman spectrum typically shows an additional peak at lower wavenumbers ~1300 cm$^{-1}$ associated with *sp$^3$* orbitals, the so-called *D* peak.[17-19] The negligible amplitude of this peak in the spectra from our samples, and the reproducibility of the Raman spectra taken from different spots in the array attest to the high quality and uniformity of the CNT grown with the ferric nitrate catalyst ink. In order to ensure the desired growth of SWNT after the DPN process, it was vital that both substrates and pen tips were properly prepared (Supporting Information SI-2). For parallel patterning, cantilever tips were dipped into specially designed inkwell chips, so that all tips could be coated with ink simultaneously (Figure 1). In principle, the inkwells can be filled with different "inks," thus increasing the flexibility of the process. After the cantilevers were coated with ink, the ink was deposited onto the surface of the substrate at the specific locations under computer control (Figure 2a and 2b). Transfer of the catalyst occurred during contact between the pen tips and the substrate surface as facilitated by formation of a water meniscus at the contact point. (For technical details see Supporting Information SI-2). Actual patterning was performed with the NanoLithography Platform (NLP) system from NanoInk, Inc. This instrument can be used with a variety of pen arrays (a single pen or 1D and 2D arrays of pens), and it is optimal for writing patterns over large areas in parallel.[15] For this study, a 1D array of 12 pens was employed for patterning. The patterns chosen were large parallel arrays of dots separated by a distance of 5 μm or 10μm. The minimum diameters of the liquid dots that were deposited on the substrate surface in our studies were in the range of 500 nm to a few microns. One could reduce this dimension considerably by controlling the ambient conditions during writing, but we found that this size gave us the desired density of CNT after CVD growth; our goal was to achieve sufficiently sparse growth so that individual CNT could be isolated and metallic contacts made to them by additional lithographic steps. A scanning electron micrograph of CNT emanating from the DPN patterned array is shown (Figure 3c), with a close up of one such particle containing three tubes (Figure 3d).

The synthesis of CNT was carried out in our home-built CVD system. Prior to the CVD synthesis, substrate with the ink deposited on its surface was dried in air and then placed into the tube furnace for heat treatment of the substrate with the Ar/$H_2$ mixture. In this study methane gas was used as the carbon source at 850-900 ºC to initiate the growth of carbon nanotubes. (More details on the CVD process are described in experimental section and Supporting Information SI-3).

After the synthesis step, substrates were examined so that specific arrays of patterned carbon nanotube structures could be identified for characterization and subsequent device fabrication.

While the diameter of the dots of ink initially patterned on the substrate is typically 500 nm to a few microns, the AFM image (Figure 2c) indicates that the resulting size of the cluster of catalyst nanoparticles that remain after heat treatment in $Ar/H_2$ atmosphere is much smaller, on the order of a few hundred nanometers. In fact, some AFM and SEM images of samples after DPN patterning and CVD growth showed three types of catalyst deposits that may result after the catalyst ink is dried and the substrate heat treated prior to the CVD growth process. In addition to preferred nucleation of catalyst particles in the center of the patterned dot as shown in Figure 2c, less desirable deposits of catalyst nanoparticles across the diameter of the dot and circular deposits of catalyst particles around the perimeter of the dot were also occasionally observed. The pattern finally obtained depends on several parameters, including composition of the ink, surface and tip properties, contact time, and the ambient conditions during the drying of the liquid ink droplet.[20-22] The probability of obtaining these other deposit formations was greatly reduced by using the optimized ink composition. Figure 4a and 4c show topographic AFM images acquired in non-contact mode of CNT grown on a silicon substrate from DPN patterned dots. The line profile measurement of the CNT taken along the red line (Figure 4c), indicates a tube diameter of 1.7 nm, consistent with the size of single-walled CNT (Figure 4d). We have also fabricated CNT on quartz substrates grown from DPN dots, where nanotubes are aligned along certain crystallographic directions of the substrate as demonstrated earlier.[5,14,23] In addition, we were also able to pattern dots via

DPN on a substrate not far from the etched-in silicon trenches resulting in growth of individual suspended carbon tubes across the gap (Figure S3). The particular molecular ink composition developed could potentially be used with other types of probes such as fountain pens with single- or multi-channeled cantilevers. When micro- or nano-channeled probes are used to deliver inks to the substrate from the ink reservoirs, it is very beneficial to use predominantly water-based inks, as opposed to colloidal suspensions of particles, to minimize and prevent clogging and/or residual build-up inside walls of the channels. Furthermore, the issue of ink evaporation must be considered carefully for successful large-area patterning over a span of several hours as described above.

The combination of the ability to pattern and grow CNT in specific locations by DPN and the optimal density of CNT produced by this procedure makes the fabrication protocol developed in this study ideal for fabricating devices that incorporate isolated CNTs. For example, using the alphabetical letters as markers fabricated by electron-beam lithography in the first step of the process, we can locate suitable isolated SWNT using either atomic force microscopy or scanning electron microscopy which saves a significant amount of time during further characterization steps of individual CNT. Subsequently, metallic electrodes can also be designed and patterned using electron beam lithography. An example of such a device is shown in SEM (Figure 4 (b)) and AFM (inset of 4(b)) images, made with a long SWNT grown with the DPN-CVD process described above.

In conclusion, Dip Pen Nanolithography provides a flexible route for creating large arrays of SWNT in precisely patterned locations, so that they can be located easily for subsequent processing and analysis. By employing multi-pen cantilever arrays for direct writing of the catalyst ink, faster production of patterned structures over larger areas on the substrate has been demonstrated. The quality and the density of the resulting SWNTs make them suitable for fabricating nanometer-scale devices. By developing an appropriate ink composition reproducibility in patterning and CVD growth of high quality CNT were assured. Here, 12-pen cantilever arrays were employed that enabled a faster parallel writing process. However, DPN instruments with hundreds of thousands of pens in parallel are already commercially available, so the method can readily be scaled up to make large arrays of patterns for the templated growth of SWNTs.

**Experimental Section**

*E-beam lithography*

Commercially obtained boron doped (p-doped) or phosphorus doped (n-doped) Si wafers (Resistivity < 0.005 Ω-cm) with 300-400 nm thick thermal oxide top layer were used as substrates. The Si wafers were diced into 1 cm x 1cm chips. An array of metallic fiducial marks in the form of pairs of alphabetical characters was then defined by e-beam lithography. Spacing between each pair was 50 µm and approximately 7µm by 3µm in lateral dimension. To pattern the substrates a standard bilayer e-beam resist (MMA/PMMA) was used. The samples were exposed in a Tescan MIRA field emission scanning electron microscope that had been adapted for use as an electron-beam lithography tool. The exposure was carried out with a beam voltage of 30 kV and probe current of 600 pA, typically for about 8 minutes for an array of 30 by 30 fiducial marks. After e-beam patterning, a thin layer of chromium was deposited on the substrate in a home-built e-gun evaporator at a typical rate of 0.01 nm/sec and a base pressure of $10^{-6}$-$10^{-7}$ torr. Prior to deposition, an *in situ* etch with 40 millitorr of $O_2$ or Ar for 12-15 seconds was performed as a cleaning step.

*Patterning and Characterization tools*

DPN patterning was done with a commercial instrument, the Nanolithography Platform (NLP 2000) manufactured by NanoInk, Inc. (this instrument has motorized stages with a resolution of ~15 nm for all three axes). 1D arrays containing 12 pens with inter-tip spacing of ~ 62 μm were employed.[27] Custom machined inkwells with matching pitch were used for coating DPN probe tips in this study[28] (Supporting Information SI-1). The imaging and analysis of the samples after the CVD growth were completed with a Scanning Electron Microscope (Nova Nano SEM 600 made by FEI), an AFM (Park XE-150), and a Renishaw InVia Raman microscope with a fixed laser excitation wavelength of 514.5 nm (2.41 eV) and fixed polarization. The SEM used for imaging was operated at 1 kV for the enhancement of the contrast (Supporting Information SI-4).

*CVD method*

After patterning by DPN, the substrates were transferred into a home-built CVD system. This system consisted of a Lindberg Blue M tube furnace with a 1" diameter quartz tube.[25] The flow rates of gases were controlled by Sierra Instruments mass flow controllers regulated by a 4-channel Sierra Instruments digital control box.[26] Substrates were placed into the CVD furnace and heated initially to 400 °C with a mixture of $Ar/H_2$ gases at flow rates of 195/175 sccm respectively for 1 hour. Prior to CNT growth the furnace was heated up to 850 °C and the flow of Ar was switched off. The temperature was stabilized for 5 minutes with only pure $H_2$ gas flowing at 150 sccm for 10 minutes. To initiate the growth of the CNT, the temperature was then raised to 900 °C and a mixture of $CH_4/Ar/H_2$ at flow rates of 900/60/140 sccm respectively was introduced. The $CH_4$ acts as the carbon source for the growth of the nanotubes. The growth time was typically 10-15 minutes, after which the flow of $CH_4$ was abruptly stopped and the furnace was allowed to cool to room temperature under the aforementioned $Ar/H_2$ atmosphere.

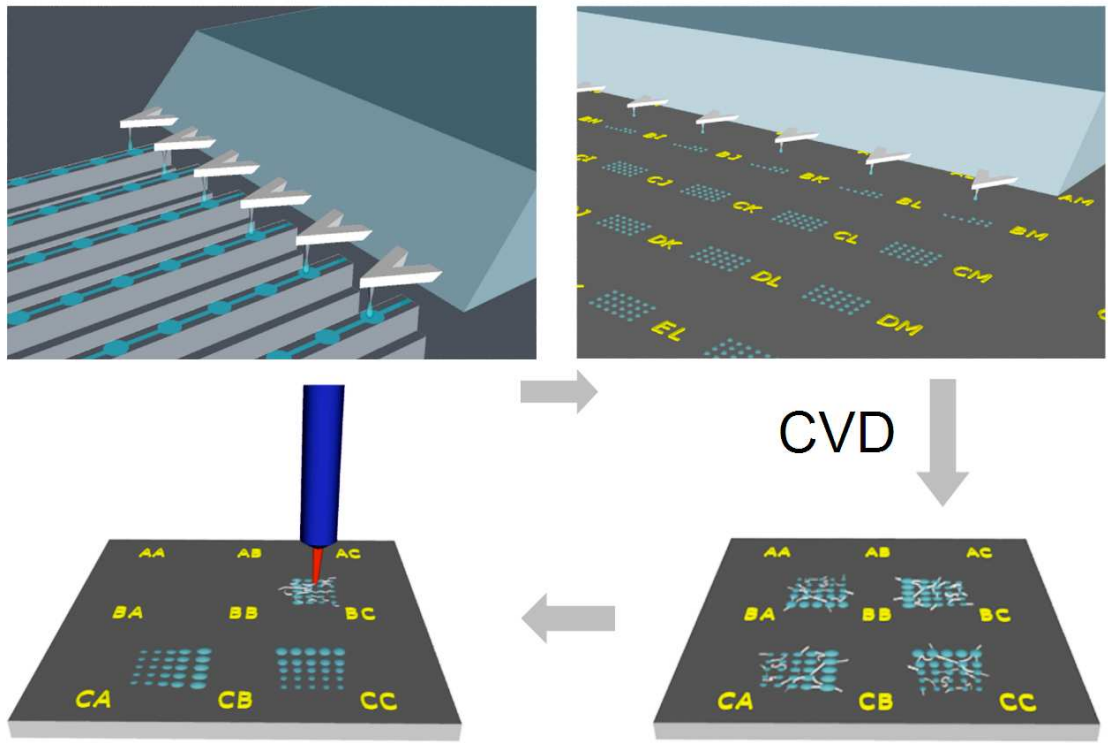

**Figure 1|** Schematic of the process of fabricating SWNT using Dip Pen Nanolithography: 1) An array of twelve pens was first dipped into reservoirs containing the iron-based solution that acts as the ``ink.'' The substrate was an oxidized silicon wafer on which alignment marks have previously been fabricated using electron-beam lithography. 2) The pattern designed on a computer (in this case an array of dots) was then written on the substrate. 3) The chip was then placed in a tube furnace for the CVD growth of the SWNT. 4) The samples were then characterized by scanning electron microscopy, atomic force microscopy and Raman spectroscopy.

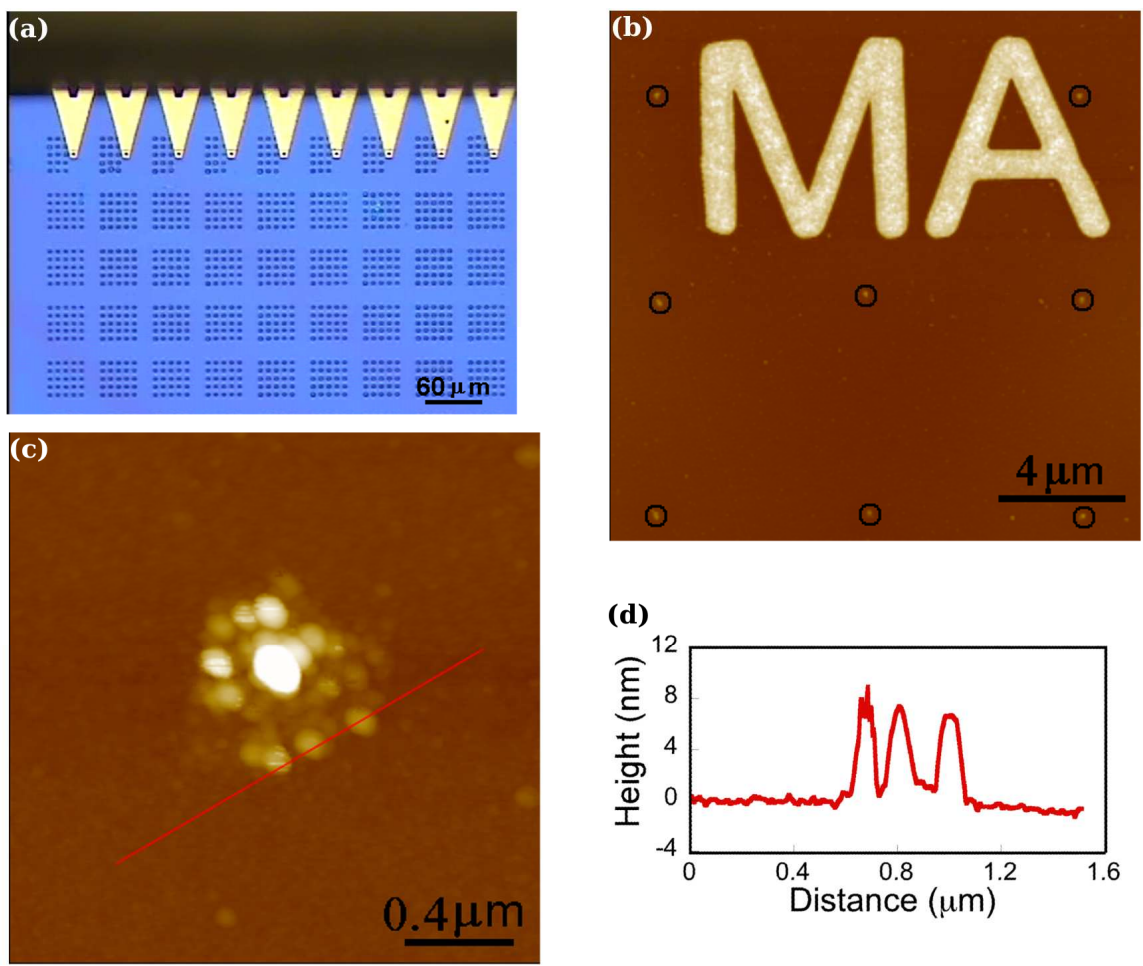

**Figure 2 |** Optical and AFM images of DPN-prepared dot arrays patterned with molecular Fe-based ink on a Si/SiO$_2$ substrate after drying and annealing; a) optical image of patterning multiple square arrays, only a 9-pens section of the 12-pen array is shown; b) AFM image of 3x3 dot array. DPN patterned dots small bright points are incircled (black) for clarity. The letters ``MA'' are Cr reference marks; c) AFM image of an individual dot from Fig. b); d) Topography profile along the line shown in c), showing the heights of individual nanoparticles inside the dot. The height of individual nanoparticles is approximately 6 nm.

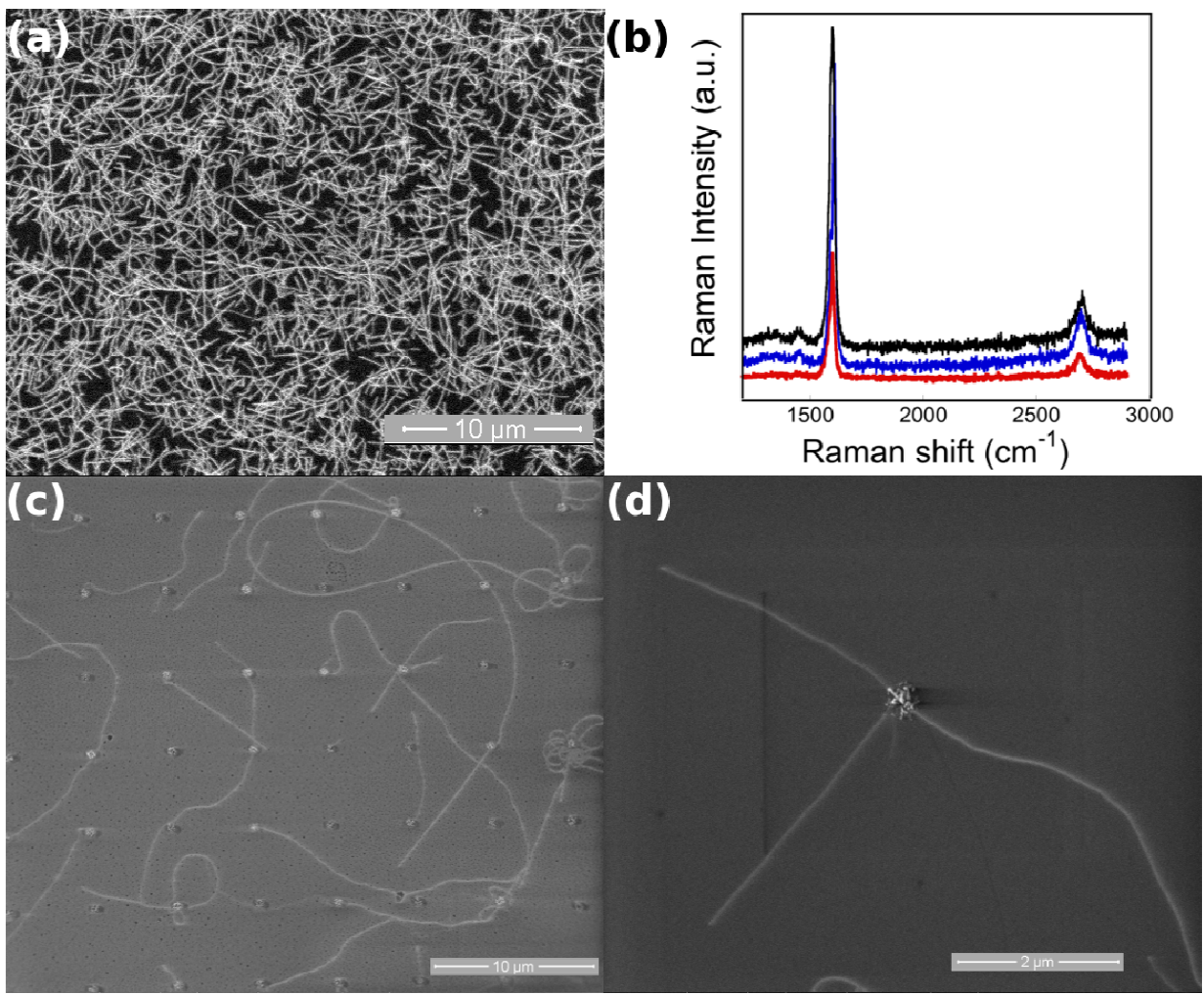

**Figure 3** | (a) Network of SWNT grown by CVD on Si/SiO$_2$. The catalyst was deposited by dipping the wafer in the same ``ink'' used in the DPN process. (b) Raman spectra taken at three random spots on the sample shown in (a). The curves are offset for clarity. (c) Scanning electron micrograph of a DPN patterned array of dots after synthesis of CNT. (d) Enlarged image of a typical dot with a few individual tubes.

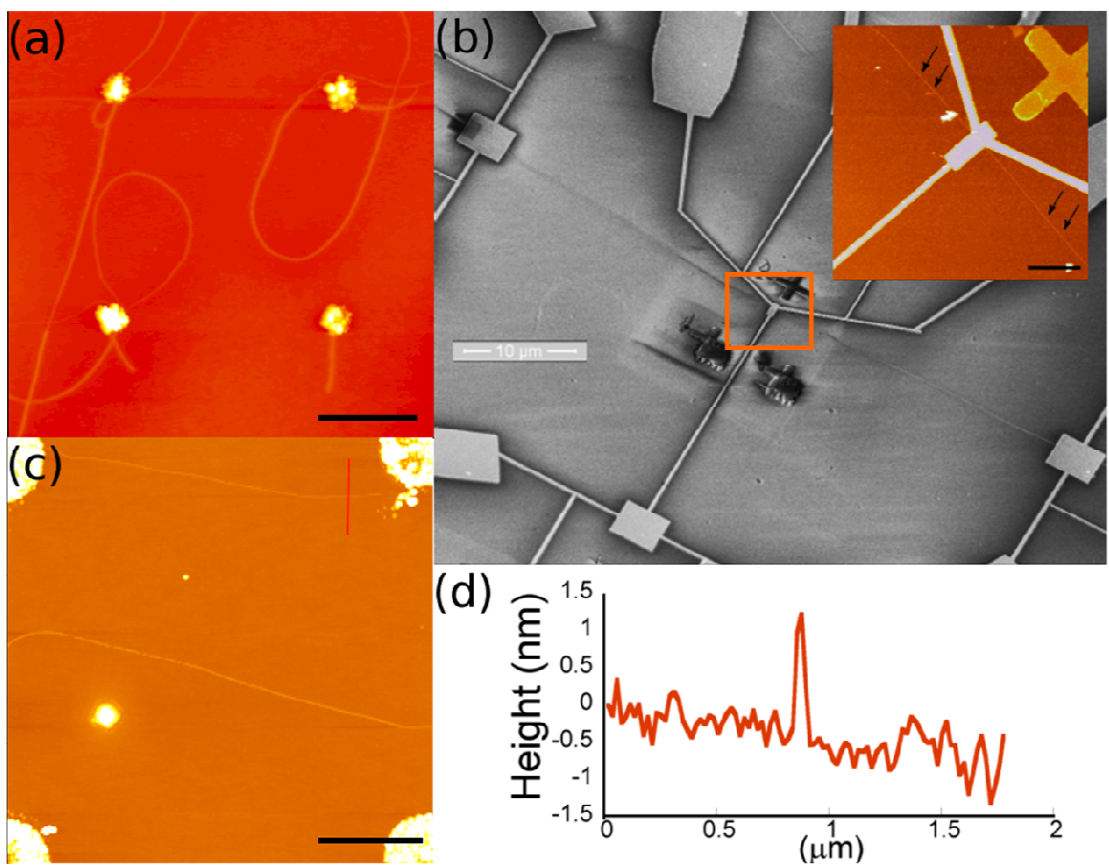

**Figure 4 |** AFM and SEM images of individual CNT and a contacted CNT with electrical leads. (a) Close up AFM image showing 4 dots patterned by DPN after CVD growth. The size bar is 2 μm. (b) Scanning electron micrograph of a long SWNT contacted with metallic electrodes; inset is AFM image of a boxed area in (b), size bar is 2 μm, arrows point to SWNT, (c) AFM image of a SWNT bridging two catalyst dots. The size bar is 2.5 μm. (d) Topographic line profile shown in red in Fig 4(c), demonstrating a tube diameter of 1.7 nm.

# Supporting Information

## SI-1 DPN 12 pen array patterning process

After deposition of the fiducial marks, but prior to patterning by DPN, the substrates were heat treated in the CVD furnace at 450 ºC for 1 hour in a flow of a Ar/H2 gas mixture at flow rates of 495 and 475 sccm respectively, in order to remove residues of e-beam resist or other organic contaminants.[1] DPN patterning was done with a commercial instrument, the Nanolithography Platform (NLP 2000) manufactured by NanoInk, Inc. This instrument has a motorized stage with a resolution of ~15 nm, and it can pattern uniform features with dimensions down to 60 nm. It is optimal for parallel writing using arrays of cantilevers; in this study a 12-pen 1D array of cantilevers was employed (Figure S1). The cantilever tips were cleaned in oxygen plasma for 20 seconds to render them hydrophilic, which ensured that the ink wetted the cantilever tips well. High-resolution optical leveling was performed to align the cantilever tips in parallel to the substrate. With a single dip into the inkwell, an array of approximately $10^4$ dots can be written by a single cantilever tip. For producing an array of dots over a large area of the substrate, the cantilever tips can be repeatedly dipped into the inkwells, allowing automated patterning of large arrays over a span of hours, taking advantage of the high resolution of the NLP instrument. Figure S1 shows the inking and patterning process. Patterning by an individual tip is quite uniform. However, occasional variations in writing performed by a particular pen are observed (Figure S1c and d). When patterning is done with multiple tips, the condition and shape of individual tips affect the resolution of the writing and the patterned dot size. Addressing each individual tip with separate force feedback control could minimize such effects. Existing attempts to employ a feedback control for an individual tip on multi-pen cantilevers has proven to be challenging.[2] A movie of the DPN writing process with the NLP is available as part of the online Supporting Information.

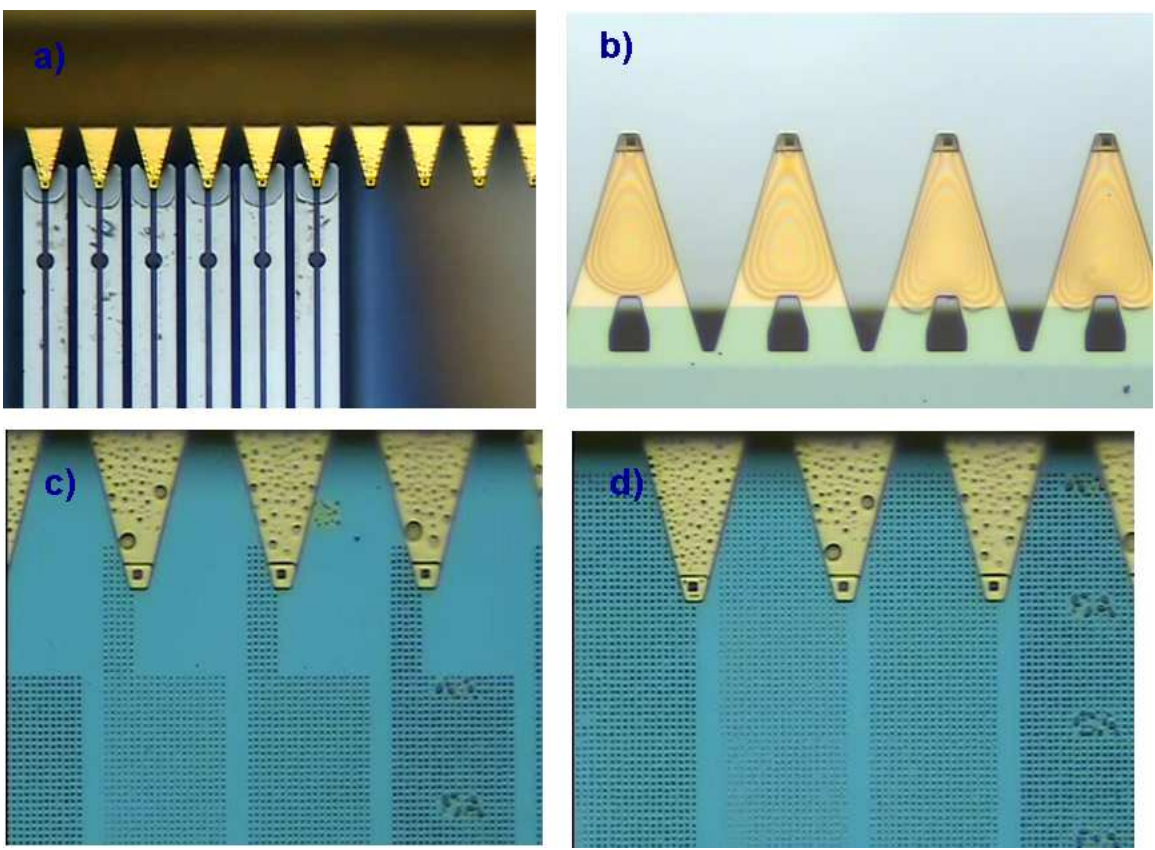

**Figure S1** | Optical images of a) 6 tips dipped in ink wells for inking, b) bottom view of 4 tips on the cantilever, c)-d ) Simultaneous patterning by a 12 pen cantilever array (only 4 pens are shown on the picture).

## SI-2 Ink composition

The most critical component of the successful DPN process is the composition of the ink. For catalysts we have tried two different salts, ferric chloride and ferric nitrate. As the results from both salts were similar, we concentrated primarily on the ferric nitrate-based inks. Preparation of the ferric nitrate based ink is described below.

The ink was prepared by mixing a water-based master solution of ferric nitrate $Fe(NO_3)_3 \cdot 9H_2O$ with dimethyl formamide (DFM) and glycerol. The master solution was prepared by dissolving 0.175-0.2 mg of the ferric nitrate salt in 1 mL of 18 MΩ-cm deionized water. This stock solution was sonicated for 20 minutes prior to mixing with other ink ingredients.

The factors that are important for DPN writing are the volatility and viscosity of the ink. Several ink compositions were tested in this study. The master solution by itself did not show reproducible patterning, due to the fast evaporation of the water solvent. Adding DMF and glycerol, which both have lower vapor pressures than water under ambient conditions, significantly improved reproducibility of the patterning and at the same time did not show signs of poisoning the catalyst during the CVD growth. Evidence of the importance of the volatility of the solvents could be seen by observing the contact angle of a drop of ink solution on the $SiO_2$ surface as a function of time. With the optimal composition noted below, we found that the contact angle of the liquid with the surface changed from 50º to 34º degrees within 5-10 minutes after the drop was deposited on the surface, with no significant change after that. These observations are consistent with the model that the water from the micron size droplet evaporates within seconds but the DMF and glycerol, which have lower vapor pressures, remain.[3,4] After multiple trials, an optimal composition of the ink (master solution, deionized water, DMF, and glycerol in a 6:2:3:1 ratio by volume) was determined. With this composition, the average patterned droplet was about 1-3 μm in diameter, corresponding to a volume of about a 1 fL of ink solution. The patterning was done at a relative humidity of 50% and a temperature of 22ºC. Dots were created with a dwell time of 0.3 seconds.

## SI-3 CNT synthesis results

In order to verify the importance of ink composition for patterning and CVD growth, different ink compositions were evaluated. Varying the ink composition resulted in different shapes of the patterned dots, and significantly lower yields of isolated carbon nanotubes. For example, Figure S2 shows the results of using two different compositions of ink: 1) a mixture of the master solution and DMF in a ratio of 9:1 by volume (Figure S2 a) and b); and 2) a mixture of 1:9 by volume of the master solution and DMF (Figure S2 c) and d). The dot size varied from one to a few microns. The dot shapes were dramatically different and the growth yield for high quality CNT was significantly reduced compared to the optimized ink composition. For the 1:9 mixture, the catalyst appeared to deposit on the perimeter of the dots, while for the 9:1 mixture, the catalyst formed a uniform film covering the entire area of the dot.  In the former case (Figure S2c and d], most of the CNT grown were short, with only a few longer CNT bridging the dots.  In the latter case (Figure S2) a and b], individual clusters of nanoparticles were difficult to obtain.  Hence the yield of CNT was significantly reduced.  In contrast, when the optimal composition of the ink was used, as described earlier, the patterned dots were filled with small clusters of catalyst nanoparticles, resulting in high-quality CNT with the desired density.  Small variations of the order of 1-3% in composition of this optimal ink did not affect the results (Figure 3).  The top panel presents SEM images of DPN dot arrays and CNT bridging from dot to dot.  The bottom panel shows SEM image where several isolated CNT originating from DPN patterned dots and extending across the gap of 5 μm in etch-in Si. Similar results were also obtained for 10 μm wide trenches with isolated tubes suspended across the gap.

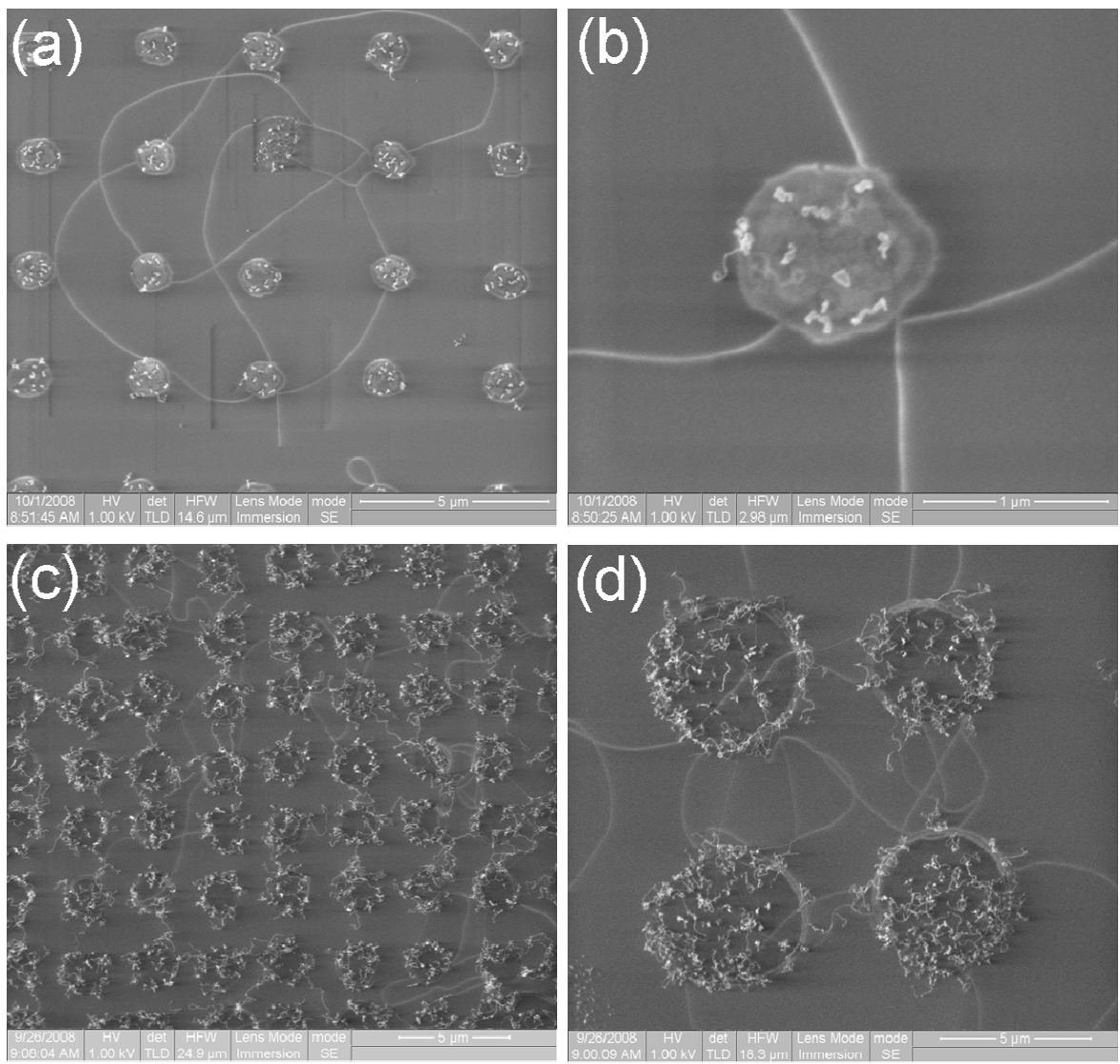

**Figure 2 |** SEM images of DPN dot arrays and CNT with different ink compositions; a)-b) a mixture of the master solution and DMF in a ratio of 9:1 by volume, and c)-d) 1:9 by volume.

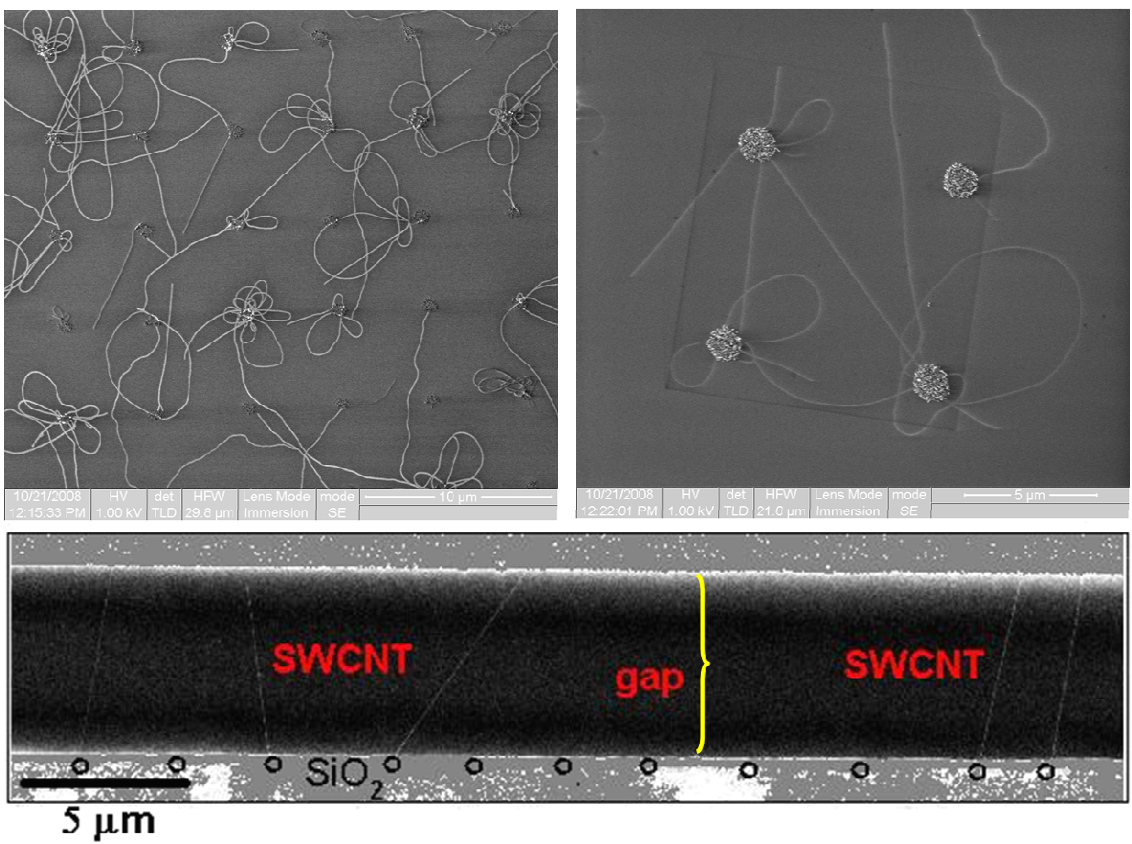

**Figure 3 |** Representative SEM images of DPN dot arrays and CNT after the CVD growth with optimal ink composition. Top pannel: 6x6 (left) and 2x2 (right) dot arrays differ only by 3 percent in ink composition. Individual nanotubes are seen to originate form each dot. Bottom pannel: several isolated CNT are shown to grow acorss the tranch in etch-in Si from pattened dots (dots are not seen in the image due to adjustment of conrasts to show CNT, black circles are added for giudance)

## SI-4 Characterization of CNTs

Thin networks of CNT grown from ink solution were examined by Raman spectroscopy. The spectral analysis was done according to the methods proposed and published by the Dresselhaus group at the MIT.[6,7] Raman spectra shown here were acquired with a laser excitation wavelength of 514.5 nm (2.41 eV) and fixed polarization. The resulting Raman spectra from the carbon nanotube network sample ( Figure S4 ) exhibit a characteristic first-order Raman *G* band double-peak, with a peak *G+* at ~1604 cm$^{-1}$ and a second peak *G-* at ~1577 cm$^{-1}$, (Figure S4a). The shape and linewidth of the peaks can be used to characterize the quality and properties of the CNT. The narrow linewidth and the Lorentzian line shape are consistent with the assumption of high quality semiconducting SWCNT. For example, the $G+$ peak at ~1604 cm$^{-1}$ has a fitted full-width at half maximum (FWHM) of ~11 cm$^{-1}$. The $G-$ peak at ~ 1577 cm$^{-1}$ is slightly broadened (FWHM) ~14 cm$^{-1}$, most likely due to interactions between tubes and the distribution in diameters of the tubes over the ~0.5-0.7 µm diameter of the Raman probe. In general, topographic AFM images on our samples show that diameter of CNT range between 1.5- 4 nm (see Figure 4d in the main text). Although the Raman results indicate the presence of semiconducting tubes of small diameter mainly contributing to the resonant Raman signal, they do not exclude the possibility of metallic tubes present that are not resonant at the fixed excitation wavelength. Note that the absence of the $D$ band peak around ~ 1300-1340 cm$^{-1}$ that is typically associated with disorder implies that all probed SWNTs are of high quality.

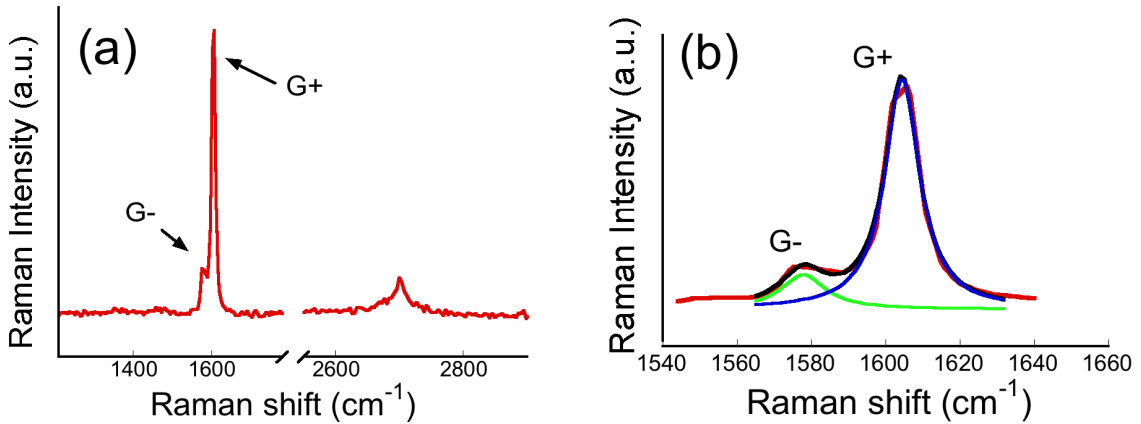

**Figure S4** | Raman spectra form a) networks of CNT grown on Si substrates from Fe-based molecular ink. Characteristic G band double peak at ~1580 cm$^{-1}$, and a small second order *2D* peak at ~2700 cm$^{-1}$ is also noticeable; b) The *G* band double peak fit with a sum of two Lorentzians for *G+* and *G-*. Green and blue curves in (b) fit individual peak to G- and G+ correspondingly, black curve is double peak Lorentzian fit to experimental data (red curve).